\documentclass[journal,12pt,onecolumn,draftclsnofoot,]{IEEEtran}

\usepackage{amsthm,amsmath,amssymb}
\usepackage{mathrsfs}
\usepackage{amsfonts}
\usepackage{graphicx}
\usepackage{booktabs}
\usepackage{algorithm}
\usepackage{algorithmic}
\usepackage{graphics}
\usepackage{subfigure}
\usepackage{multirow}
\usepackage{titlesec}
\usepackage{color}

\ifCLASSINFOpdf

\else

\fi

\usepackage{amsmath}
\begin{document}
\title{\Large{Data Augmentation Empowered Neural Precoding for Multiuser MIMO with MMSE Model }}

\author{\normalsize{Shaoqing Zhang, Jindan Xu,~\IEEEmembership{\normalsize{Member,~IEEE,}} Wei Xu,~\IEEEmembership{\normalsize{Senior Member,~IEEE,}} Ning Wang,~\IEEEmembership{\normalsize{Member,~IEEE,}} Derrick Wing Kwan Ng,~\IEEEmembership{\normalsize{Fellow,~IEEE,}} and Xiaohu You,~\IEEEmembership{\normalsize{Fellow,~IEEE}}\vspace{-1cm}}
\thanks{
This work was supported by the NSFC under grants 62022026, 61871109, and 61771431, and the Natural Science Foundation of Jiangsu Province for Distinguished Young Scholars under grant BK20190012. \emph{(Corresponding authors: Jindan Xu; Wei Xu.)}

S. Zhang, J. Xu, W. Xu, and X. You are with the National Mobile Communications Research Laboratory, Southeast University, Nanjing 210096, China. W. Xu is also with Henan Joint International Research Laboratory of Intelligent Networking and Data Analysis, Zhengzhou University, Zhengzhou, 450001 China (e-mail:\{sq\_zhang, jdxu, wxu, xhyou\}@seu.edu.cn).
}
\thanks{N. Wang is with the Department of Electrical Engineering, Zhengzhou University, Zhengzhou, Henan 450001, China (e-mail: ienwang@zzu.edu.cn).}
\thanks{D. W. K. Ng is with the School of Electrical Engineering and Telecom- munications, University of New South Wales, Sydney, NSW 2052, Australia (e-mail: w.k.ng@unsw.edu.au).}
}
\maketitle
\begin{abstract}
Precoding design exploiting deep learning methods has been widely studied for multiuser multiple-input multiple-output (MU-MIMO) systems. However, conventional neural precoding design applies black-box-based neural networks which are less interpretable. In this paper, we propose a deep learning-based precoding method based on an interpretable design of a neural precoding network, namely iPNet. In particular, the iPNet mimics the classic minimum mean-squared error (MMSE) precoding and approximates the matrix inversion in the design of the neural network architecture. Specifically, the proposed iPNet consists of a model-driven component network, responsible for augmenting the input channel state information (CSI), and a data-driven sub-network, responsible for precoding calculation from this augmented CSI. The latter data-driven module is explicitly interpreted as an unsupervised learner of the MMSE precoder. Simulation results show that by exploiting the augmented CSI, the proposed iPNet achieves noticeable performance gain over existing black-box designs and also exhibits enhanced generalizability against CSI mismatches.
\end{abstract}


\IEEEpeerreviewmaketitle

\section{Introduction}

\IEEEPARstart
{I}{n} the last two decades, the technology of multiuser multiple-input multiple-output (MU-MIMO) has received extensive attentions in wireless communications thanks to its great potential of improving the network spectral efficiency (SE). To unlock the potential of MIMO communications, precoding designs have been intensively studied, e.g., \cite{3}, \cite{2}. 

In general, due to the nonconvexity of multiuser precoding optimization problems, it is nontrivial to find the optimal precoding design for SE maximization in MU-MIMO networks. For implementation simplicity, three types of linear precoding methods, i.e., maximum ratio transmission (MRT) precoding, zero-forcing (ZF) precoding, and minimum mean-squared error (MMSE) precoding, were introduced with closed-form solutions offering important system design insights \cite{2}. To further improve the system performance, a series of efficient precoding design methods by applying conventional convex optimization tools have been proposed to address the design problem. For instance, an iterative transceiver optimization method, named weighted minimum mean-squared error (WMMSE), was proposed in \cite{3} for an interference-limited broadcast channel. In general, the iterative precoding method usually costs much run time to obtain precoding matrix in online deployment.

Recently, deep learning (DL) methods, due to their strong capability of function approximation, have triggered evolutions in re-designing physical-layer technologies, e.g., channel estimation \cite{5}, channel state information (CSI) compression \cite{6}, \cite{7}, and signal detection \cite{8}. In particular, the application of DL to efficient precoding designs under various scenarios has been attracted much attention \cite{10}. The main idea of DL-based precoding design is to convert online calculations of a traditional algorithm into an offline training process by using deep neural networks (DNNs), e.g., fully-connected neural networks (FNNs) and convolution neural networks (CNNs). Indeed, a series of studies have shown that precoding design using DL can even achieve better performance compared with the commonly adopted optimization methods, while significantly reducing the complexity of online computation \cite{11}.
\begin{figure*}[t]
    \centering
    \includegraphics[scale=0.5]{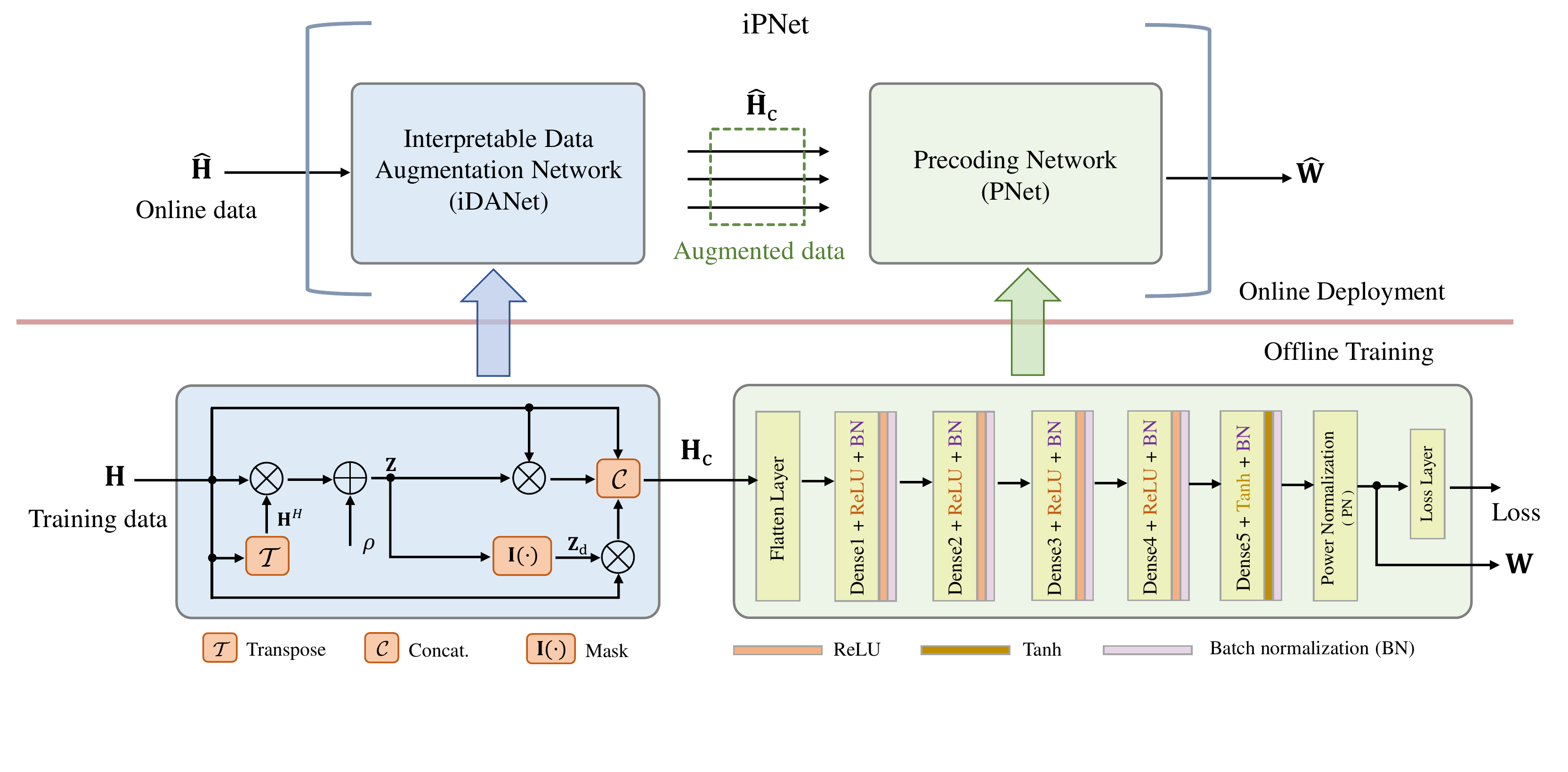}
    \caption{A two-stage block diagram of the proposed iPNet.}
\end{figure*}

Existing DL-based precoding methods follow two popular approaches: black-box neural network (NN) design and deep-unfolding network design. Specifically, black-box NN methods replace the precoding optimization algorithm by a well-trained black-box NN, which learns the mapping relationship between the input CSI and the output precoding matrix. For example, in \cite{10}, a framework of CNN was proposed for the digital beamforming design in a multiple-input single-output (MISO) system. Also, in \cite{11}, a DL-based precoder was proposed for fast beamforming calculations. Besides, a simple black-box FNN-based design was studied in \cite{9} for optimizing the beamforming matrix in a single-user MISO system by adopting a two-stage precoding approach. These NN-based methods learned the mapping from perfect knowledge of CSI to the desired beamforming matrix. On the other hand, by considering a more practical assumption of imperfect CSI, an FNN-based approach was recently proposed in \cite{12}. However, these black-box NN designs are hardly interpretable, thus lacking the generalizability against various imperfections in practice. To achieve interpretability, a deep-unfolding NN was proposed in \cite{44}, which unfolded the iterations of a conventional optimization algorithm into a layer-wise structure of the proposed NN. Compared to black-box NNs, the deep-unfolding NN offers a better interpretability, which guarantees its performance. However, the design of this deep-unfolding NN is computationally expensive and sometimes requires specifically derived back propagation expression, which is less extensible to different scenarios [11]. 

In this paper, in order to facilitate an efficient and concise design by tackling the challenges of network interpretability and generalizability, we investigate the DL-based precoding design in an MU-MIMO system and develop an MMSE model-driven interpretable neural precoding network, namely iPNet, for optimizing the precoder for SE maximization. The proposed iPNet is a hybrid network composed of a model-driven component network responsible for input CSI augmentation and a data-driven sub-network for the subsequent precoder calculation from the augmented CSI. Compared to the traditional MMSE precoding, the proposed iPNet achieves better performance in terms of sum rate and exhibits less computational complexity.
Besides, compared to the existing data-driven black-box DL-based methods, e.g., [9], the iPNet is more interpretable for system performance improvement, while offering enhanced generalizability against CSI mismatches.


\section{System Model}
We consider an MU-MIMO system with $M$ antennas equipped at the base station (BS) serving 
$K$ single-antenna users.
The received signal at user $k$, $\forall k\in \left\{ 1,\cdots ,K \right\}$, is expressed as
\begin{equation}
\mathrm{y}_k=\mathbf{h}_{k}^{H}\mathbf{w}_ks_k+\sum_{j=1,j\ne k}^K{\mathbf{h}_{k}^{H}\mathbf{w}_js_j}+n_k
\triangleq \mathbf{h}_{k}^{H}\mathbf{Wd}+n_k,
\end{equation}
where ${\mathbf{h}_{k} \in \,}$$\mathbb{C}$${^{M\times 1}}$ is the complex-valued downlink channel from the BS to user $k$, ${\mathbf{W}=}$ $[ \mathbf{w}_1,\mathbf{w}_2,$ $\cdots ,\mathbf{w}_K] \in \,$ $\mathbb{C}$ ${^{M\times K}}$ is the downlink precoding matrix with $\mathbf{w}_k \in \,$$\mathbb{C}$${^{M\times 1}}$ denoting the precoding vector for user $k$, ${n_{k}\sim \mathcal{C} \mathcal{N} \left( 0,\mathrm{\sigma}^2 \right) }$ is the additive white Gaussian noise (AWGN) with zero mean and variance $\mathrm{\sigma}^2$ at user $k$, $s_{k} \in \,\mathbb{C}$ is the transmitted data symbol with normalized average symbol energy for user $k$, i.e., $\mathrm{E}{\left\{ \left| s_k \right|^2 \right\}}=1$, and $\mathbf{d}=\left[ s_1,...,s_K \right] ^T\in \mathbb{C} ^{K\times 1}$ is the transmitted signal vector at the BS. Note that for single-user MIMO system, the receiving data at all antennas can be jointly detected at this single user. Different from the single-user system, the MU-MIMO system considered is more practical and the receiving data is detected by the corresponding users separately and multiuser interference presents.

For the system described by (1), our objective is to maximize the downlink sum rate of all users by optimizing the precoding matrix $\mathbf{W}$. It can be formulated as
\begin{equation}
  \begin{aligned}
    &\underset{{\mathbf{w}_k}}{\mathrm{maximize}} \,\, R\overset{\bigtriangleup}{=}\sum_{k=1}^K{\log _2\left( 1+\frac{|\mathbf{h}_{k}^{H}\mathbf{w}_k|^2}{\sum_{j=1,j\ne k}^K{|\mathbf{h}_{k}^{H}\mathbf{w}_j}|^2+\sigma^{2}} \right)} \\
    & \mathrm{subject\,to}\,\,\,\,\,\,\,\,\,\,\,\,\, \sum_{k=1}^K{\left\| \mathbf{w}_k \right\| ^2}\leqslant P_{\mathrm{T}},
\end{aligned}  
\end{equation}
where $P_{\mathrm{T}}$ is the total maximum transmit power budget. Note that the sum rate maximization considered is a popular optimized objective seen in much literature, e.g., [7]-[11], and using the sum power constraint in (2) is motivated in practice by the limited budget of the BS.

For simplicity, the optimization in (2) is regarded as finding a mapping from the CSI, $\mathbf{H}$, to $\mathbf{W}$, denoted by a mapping function as $\mathcal{W} : \mathbf{H}\rightarrow \mathbf{W}$, where $ \mathbf{H}=\left[ \mathbf{h}_{1}^{H},\mathbf{h}_{2}^{H},\cdots,\mathbf{h}_{K}^{H} \right]\in \mathbb{C} ^{M\times K}$ is the stacked channel of all the users.

The mapping, $\mathbf{W}=\mathcal{W}\left( \mathbf{H} \right)$, is generally nonconvex and requires much effort to solve. Typically, a linear precoding scheme based on the criterion of MMSE [1] is adopted. The MMSE precoding is given by
\begin{equation}
    \mathbf{W}=\mathbf{H}\left( \mathbf{H}^H\mathbf{H}+\rho \mathbf{I}_K \right) ^{-1},
\end{equation}
where $\rho$ is the reciprocal of the signal-to-noise ratio (SNR). 
Note that this linear MMSE precoding scheme is in a closed-form but generally suboptimal.

In order to obtain $\textbf{W}$ efficiently, DL-based methods have been studied for the precoding design, e.g., straightforward black-box designs of NNs. The design of black-box NN maps directly from $\mathbf{H}$ to $\mathbf{W}$, while the network structure design usually lacks interpretability. In the next section, we introduce a NN design with input data augmentation to address these challenges faced by the black-box NN-based methods.

\section{Data Augmentation Enabled Interpretable Neural Precoding Network}
In this section, we elaborate the proposed architecture of iPNet for learning the precoding mapping, $\mathcal{W}\left( \mathbf{H} \right)$, as depicted in Fig. 1. The iPNet consists of two modules: Interpretable data augmentation sub-network (iDANet) and precoding sub-network (PNet). The architecture of iDANet is inspired by the analytical model of the linear MMSE precoding in (3), which is a model-driven module. It realizes ``data augmentation'' of the CSI input, i.e., the input CSI is preprocessed by the iDANet to establish an augmented version of the CSI with a larger dimension containing richer information. Following iDANet is the subsequent module of PNet, which is composed of a multi-layer DNN and learns the mapping from the augmented CSI transformation (output of the iDANet) to the desired precoding matrix $\mathbf{W}$. 
For training the proposed iPNet, we adopt a two-stage procedure including offline training and online deployment as shown in Fig. 1.

\subsection{Structure Design of iDANet}
The iDANet is a model-driven network without the need of any gradient update. The structure design of iDANet is guided by the analytical model of an MMSE precoder. For a typical DNN, learning the operation of inverting of a matrix is much more difficult than that of learning an approximately linear mapping, as the learning accuracy of a matrix inversion is hard to guarantee [11]. Therefore, in order to efficiently improve the learning ability of the proposed iPNet, we first convert the matrix inversion operation in (3) into a linear form by introducing a transformed and augmented CSI input.

To start with, for an inversible square matrix $\mathbf{Z} \in \,$$\mathbb{C}$${^{K\times K}}$, we apply the first-order Taylor expansion of $\mathbf{Z}^{-1}$ at $\mathbf{Z}_0$, which yields \cite{13} 
\begin{equation}
    \mathbf{Z}^{-1}=2\mathbf{Z}_0^{-1}-\mathbf{ZZ}_0^{-2}+o\left( \mathbf{Z}_{0}^{-2} \right),
\end{equation}
where $o\left( \mathbf{Z}_{0}^{-2} \right)$ is a higher-order infinitesimal of $\mathbf{Z}_{0}^{-2}$. For simplicity, we choose $\mathbf{Z}_0$ as a diagonal matrix $\mathbf{Z}_\mathrm{d}$ that only retains the diagonal elements of $\mathbf{Z}$. Note that $\mathbf{Z}_{0}^{-1}$ can be written as the linear expression $\mathbf{Z}_{0}^{-1}=\mathbf{Z}_{\mathrm{d}}\times \mathbf{Z}_{\mathrm{d}}^{-2}$, which can be regarded as a linear expression of $\mathbf{Z}_{\mathrm{d}}$, thus the term $2\mathbf{Z}_0^{-1}$ in (4) can be approximated by the structure $\mathbf{Z}_\mathrm{d}\mathbf{X}_0$, where $\mathbf{X}_0$ is a trainable matrix-form parameter with the same size as $\mathbf{Z}$. The last term $-\mathbf{ZZ}_0^{-2}+o\left( \mathbf{Z}_{0}^{-2} \right)$ in (4) can be regarded as an affine function of $\mathbf{Z}$ and we represent it as $\mathbf{ZX}_1+\mathbf{X}_2$ for approximation, where $\mathbf{X}_1$ and $\mathbf{X}_2$ are both trainable matrix-form parameters with the same size as $\mathbf{Z}$. In particular, $\mathbf{X}_2$ is introduced to match the higher-order term in (4), i.e., we expect to train $\mathbf{X}_2$ for approximating the residual term $o\left( \mathbf{Z}_{0}^{-2} \right)$, to reduce the higher-order error. As a result, $\mathbf{Z}^{-1}$ in (4) can be re-expressed as
\begin{equation}
    \mathbf{Z}^{-1}\approx \mathbf{Z}_\mathrm{d}\mathbf{X}_0+\mathbf{ZX}_1+\mathbf{X}_2.
\end{equation}

Let $\mathbf{Z}\triangleq \mathbf{H}^H\mathbf{H}+\rho \mathbf{I}_K$ and substitute (5) into (3). We have
\begin{equation}
    \mathbf{W}=\mathbf{H}\mathbf{Z}^{-1}=\mathbf{H}\mathbf{Z}_\mathrm{d}\mathbf{X}_0+\mathbf{H}\mathbf{Z{X}}_1+\mathbf{H}\mathbf{X}_2+\mathbf{X}_3,
\end{equation}
where $\mathbf{X}_3$ is a further trainable matrix-form parameter with the same size as $\mathbf{Z}$ to improve the accuracy of the approximation in (6). From (6), we transform the nonlinear mapping $\mathbf{W}=\mathcal{W}\left( \mathbf{H} \right)$ into a linear mapping. This linear mapping is denoted by
\begin{equation}
\mathbf{W}=\mathcal{\hat{W}}\left(\mathbf{H}\mathbf{Z}_\mathrm{d}, \mathbf{H}\mathbf{Z},\mathbf{H} \right),
\end{equation}
and it is expected to be learned effectively by a DNN.

It should be noted that the update from $\mathcal{W}\left( \cdot \right)$ to $\mathcal{\hat{W}}\left( \cdot \right)$ leads to some changes of the subsequent precoding training for PNet. Specifically, the PNet now requires to replace the input from a single CSI matrix, $\mathbf{H}$, to an augmented CSI transformation
\begin{equation}
\mathbf{H}_{\mathrm{c}}=\left[\mathbf{H}\mathbf{Z}_\mathrm{d}, \mathbf{H}\mathbf{Z},\mathbf{H} \right] \in \mathbb{C} ^{3M \times K}.
\end{equation}
Analogous to data augmentation in the field of computer vision \cite{shorten2019survey}, the augmented CSI data $\mathbf{H}_{\mathrm{c}}$ inspired by (7) represents the original CSI data $\mathbf{H}$ in a more comprehensive way, which facilitates the learning of the subsequent precoding sub-network. 

\begin{table}[]
\centering
\setlength{\abovecaptionskip}{0cm}
\setlength{\belowcaptionskip}{-2cm}
\scriptsize
\setlength{\tabcolsep}{10mm}{
\caption{Structure Details of PNet}
\begin{tabular}{@{}cccc@{}}
\toprule
\textbf{}                                                                                & \multicolumn{1}{c}{\textbf{Layer}}                    & \multicolumn{1}{c}{\textbf{Output dim.}}               & \multicolumn{1}{c}{\textbf{Activation}}                  \\ \midrule
\multicolumn{1}{c|}{\textbf{Input}}                                                      & InputLayer                                                 & 6 $MK$                                                & \multicolumn{1}{c}{N/A}                                    \\ \midrule
\multicolumn{1}{c|}{\multirow{4}{*}{\textbf{Hidden}}}                                    & Dense 1                                                    & 64 $MK$                                                &\multicolumn{1}{c}{BN + ReLU}                                              \\
\multicolumn{1}{c|}{}                                                                    & Dense 2                                                    & 32 $MK$                                               &\multicolumn{1}{c}{BN + ReLU}                                                \\
\multicolumn{1}{c|}{}                                                                    & Dense 3                                                    & 16 $MK$                                                 &\multicolumn{1}{c}{BN + ReLU}                                                 \\
\multicolumn{1}{c|}{}                                                                    & Dense 4                                                    & 8 $MK$                                                &\multicolumn{1}{c}{BN + ReLU}                                                 \\ \midrule
\multicolumn{1}{c|}{\multirow{2}{*}{\textbf{\begin{tabular}[c]{@{}c@{}}Precoding\\ Output\end{tabular}}}}                                    & Dense 5                                                    & 2 $MK$                                                &\multicolumn{1}{c}{BN + Tanh}                                              \\
\multicolumn{1}{c|}{}                                                                    & PN                                                    & 2 $MK$                                    
     &\multicolumn{1}{c}{N/A} \\ \bottomrule
\end{tabular}}
\end{table}


The structure details of iDANet, elaborated in the lower-half of Fig. 1, is driven by (7), where in this figure $\mathcal{T}$ is the matrix transpose operation, $\mathbf{I}\left( \cdot \right) $ is the Hadamard product with an identity matrix, and $\mathcal{C}$ denotes the operation of data concatenation. Note that the MMSE precoding always outperforms the ZF precoding in terms of sum rate, while these two precoding methods share the same complexity order. Based on this consideration, we therefore choose the MMSE precoding in our work.

\subsection{Mapping Design of PNet}
The second component network of iPNet, i.e., PNnet, realizes the calculation of a precoding matrix from the input of the augmented CSI transformation, $\mathbf{H}_{\mathrm{c}}$, and outputs the precoding matrix. It should be noted that NNs just support vector input, thereby vectorization operation $\left[ \mathcal{R} \left( \mathbf{H}_{\mathrm{c}} \right) ,\mathcal{I} \left( \mathbf{H}_{\mathrm{c}} \right) \right]$ is carried out when $\mathbf{H}_{\mathrm{c}}$ is input to PNet. Unlike iDANet, PNet is a data-driven network of DNN adopting the gradient update by applying unsupervised learning. In our design, we adopt an FNN to realize PNet. The structure of PNet is shown in Fig. 1 and the design details are summarized in Table I. In particular, layers of batch normalization (BN) and activation functions are introduced to the network to accelerate the convergence and to improve the learning performance. Note that the activation function of the output layer is Tanh, while the activation functions of the hidden layers are all ReLU. The power normalization (PN) layer corresponds to the operation of $\sqrt{P_{\mathrm{T}}}\frac{\mathbf{W}}{\left\| \mathbf{W} \right\| _F}$, which guarantees the power budget constraint. Since the goal of PNet is to design the precoding matrix that maximizes the sum rate, we design the loss layer to examine the sum rate of the MU-MIMO. 
The loss function of PNet is defined as
\begin{equation}
    \mathrm{Loss}=-\frac{1}{T}\sum_{i=1}^{T}{\sum_{k=1}^K{\log_2 \left( 1+\frac{|\mathbf{h}_{k}^{H}[i]\mathbf{w}_k[i]|^2}{\sum_{j=1,j\ne k}^K{|\mathbf{h}_{k}^{H}[i]\mathbf{w}_j[i]}|^2+\sigma^{2}} \right)}},
\end{equation}
where $T$ represents the total number of training samples and the subscript $[i]$ denotes the $i$-th training sample.

Let $\Gamma=\left\{ \mathbf{V}_l\in \mathbb{C} ^{f_l\times f_{l-1}} ,\mathbf{b}_l\in \mathbb{C} ^{f_l\times 1} \right\} _{l=1,2,3,4,5}$ be the set of parameters collecting the weight matrices and bias vectors of PNet, respectively, where $f_l$ denotes the number of neurons in $l$-th layer. Then, the operation of PNet is explicitly expressed as
\begin{equation}
\mathbf{W}=\mathbf{V}_{5}\left( \mathbf{V}_{4}\cdots \left( \mathbf{V}_1\mathbf{H}_{\mathrm{c}}+\mathbf{b}_1 \right) \cdots +\mathbf{b}_{4} \right) +\mathbf{b}_{5},
\end{equation}
where $\mathbf{W}$ is the output of PNet, and the precoding matrix is finally obtained by the PN layer to guarantee the power constraint. Note that the MMSE precoding is designed to suppress the multiuser interference via minimizing the mean squared error (MSE) of all user signals. For the proposed iPNet, it learns to mimic the MMSE precoding by using a NN with a model-driven module before a data-driven subnetwork. In this way, the multiuser interference is suppressed by the iPNet similarly as the conventional MMSE precoding does.

\begin{table}[t]
\begin{center}
\setlength{\abovecaptionskip}{0cm}
\setlength{\belowcaptionskip}{-2cm}
\caption{Number of Parameters}
\begin{tabular}{@{}c|c|c|c@{}}
\toprule
 & Total params. & Trainable params. & Non-trainable params. \\ \midrule
Black-box NN [9] & 734,752 & 730,848  & 3,904\\ \midrule
Proposed iPNet & 800,544 & 796,512 & 4,032\\ \midrule
Proposed iPNet-half & 228,448 & 226,336 & 2,112\\ \bottomrule
\end{tabular}  
\end{center}
\end{table}

\subsection{Interpretability of iPNet}
In the proposed iPNet, the iDANet is responsible for the mapping which transforms the raw CSI matrix, $\mathbf{H}$, to the augmented transformation of the CSI matrix, $\mathbf{H}_{\mathrm{c}}$, obtained by a model-driven approach. Due to the augmentation of the learning data by the iDANet, the subsequent precoding network can learn a mapping more effectively and can outperform black-box NN for poor environment, e.g., low SNR scenarios. The subsequent PNet is responsible for learning the precoding matrix directly from the augmented input of $\mathbf{H}_{\mathrm{c}}$, rather than the raw $\mathbf{H}$, which facilitates the learning of the data-driven network. The mathematical expression of (6) shares the same relationship as that of (10), which allows the PNet to learn the cascaded linear mapping from $\mathbf{H}_{\mathrm{c}}$ to $\mathbf{W}$ more effectively. 
By applying unsupervised learning with the objective of sum rate maximization, the proposed iPNet in fact outperforms the MMSE precoding, which will also be verified by the following simulations. Note that the MMSE precoding is often incorporated with some multiuser scheduling strategies for ensuring the notion of fairness. Therefore, most of the existing multiuser scheduling methods, including the proportional fairness scheduling, can be directly applied to the proposed precoding if multiuser fairness is needed.

\section{Numerical Results}

\subsection{Simulation Setups}
We test a multiuser system with $M=4$ and $K=4$. The SNR in simulation figures corresponds to the transmit power over the noise variance which is normalized to unity. We generate a total of 100,000 samples of $\mathbf{H}$ and the corresponding estimated channel $\hat{\mathbf{H}}$ is obtained by using the linear MMSE estimator. 
The Adam optimizer is adopted in the training phase [9]. 
The learning rate is initialized to $\eta =0.01$ and then reduced by one-tenth if the validation performance of three consecutive epochs drops or remains unchanged.


\subsection{Performance Evaluation} 
For fair comparison, we set the number of layers and neurons of the black-box NN to be the same as that of our PNet, and the design of BN and activation function layers of the black-box NN follow that in [9], which can rule out the influence of the variability of the numbers of layers and neurons. Besides, we also directly halve the number of neurons in the hidden layer of iPNet, namely iPNet-half, to evaluate the performance of iPNet under a smaller number of network parameters. The comparison of network parameters is listed in Table II.

Fig. 2 compares the sum rate of the MMSE precoding, black-box NN, iPNet, and iPNet-half. Notably, the iPNet and iPNet-half outperform both the MMSE precoding schemes under the cases of perfect and imperfect CSI, especially in high pilot-to-noise ratio (PNR) scenarios. This is due to the interpretable structure of iPNet and the unsupervised learning applying the loss function in (9) that is directly related to the sum rate. Besides, we observe that both iPNet and iPNet-half achieve higher performance than the black-box NN and the performance gap between iPNet and the black-box NN increases with PNR and SNR. Particularly, the proposed iPNet-half achieves better performance than that of the black-box NN while reducing the network parameters by nearly two-thirds. This is because learning an interpretable linear mapping from the augmented CSI data by the proposed iPNet is more effective than learning a nonlinear mapping from the raw CSI data by the black-box NN.

\begin{figure}[t]

\centering
\subfigure[]{
\begin{minipage}[t]{0.5\linewidth}
\centering
\includegraphics[scale=0.6]{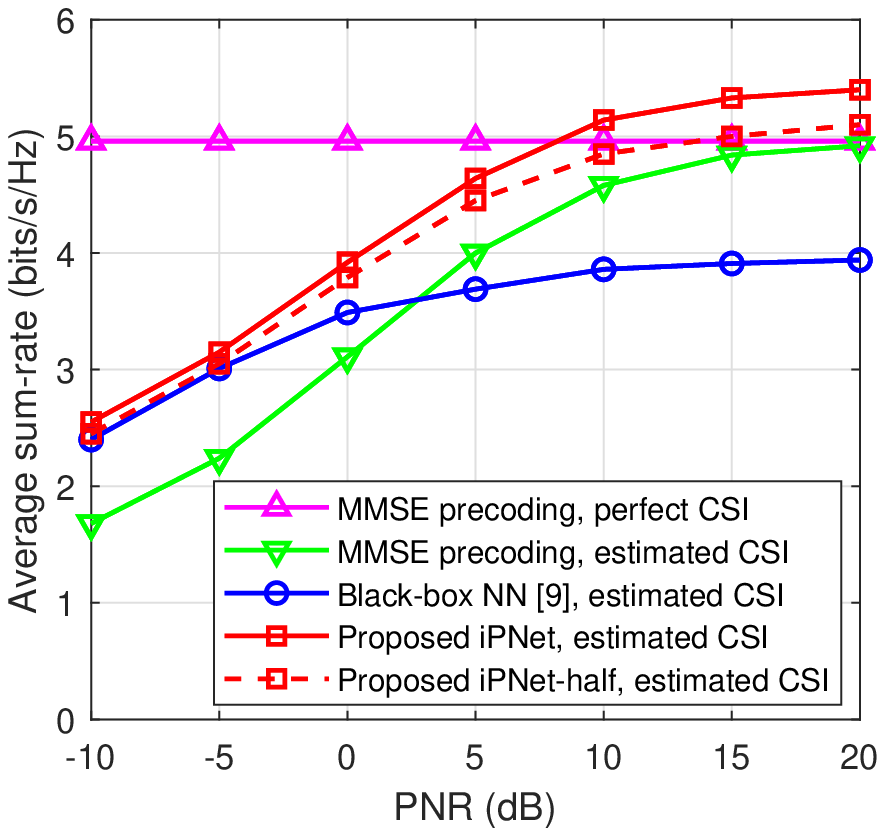}
\end{minipage}%
}
\subfigure[]{
\begin{minipage}[t]{0.5\linewidth}
\centering
\includegraphics[scale=0.6]{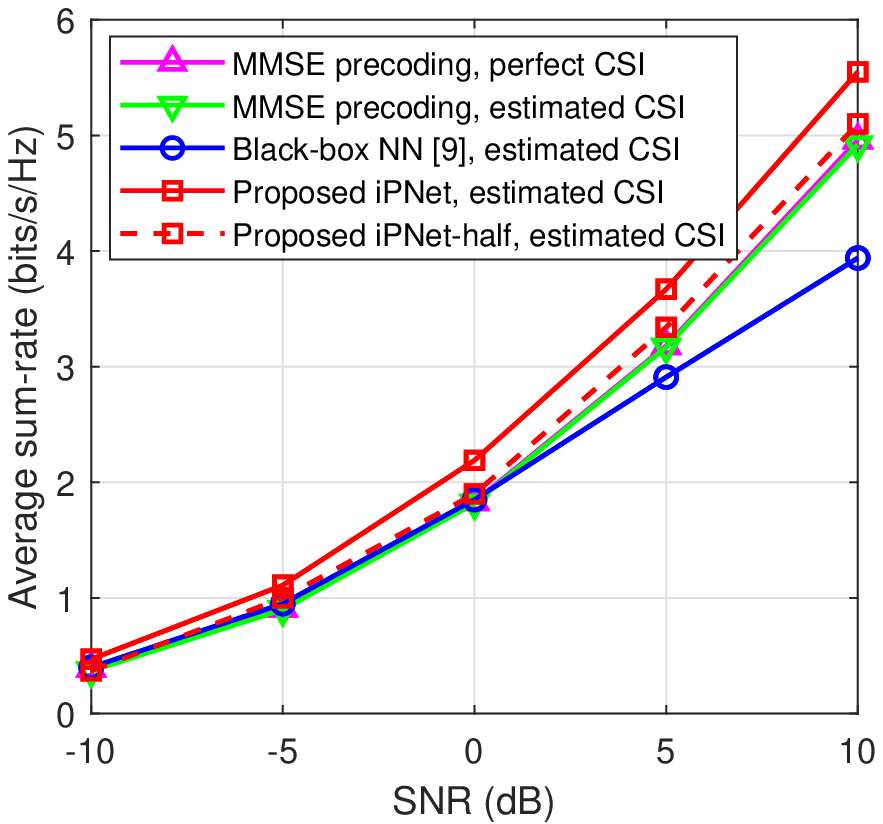}
\end{minipage}%
}%
\centering
\caption{(a). Comparison of sum rate versus PNR. (b). Comparison of sum rate versus SNR.}

\end{figure}

On the other hand, concerning the complexity comparison, the MMSE precoding requires matrix inversions and matrix multiplications and the total computational complexity is in the order of $\mathcal{O} \left( K^3+MK^2 \right)$. In contrast, the iPNet requires only ordinary matrix multiplications, and its computational complexity is in the order of $\mathcal{O} \left( MK^2 \right)$. Alternatively, in terms of the run time of precoding calculation, the trained iPNet requires 18.19 $\mathrm{\mu s}$ while the MMSE algorithm needs 22.66 $\mathrm{\mu s}$. Similar to the other DL-based precoding methods, e.g., [7]-[9], although the complexity reduction is slight, the computations of DL networks can be implemented in parallel efficiently.

\begin{figure}[t]
\centering
\subfigure[]{
\begin{minipage}[t]{0.5\linewidth}
\centering
\includegraphics[scale=0.6]{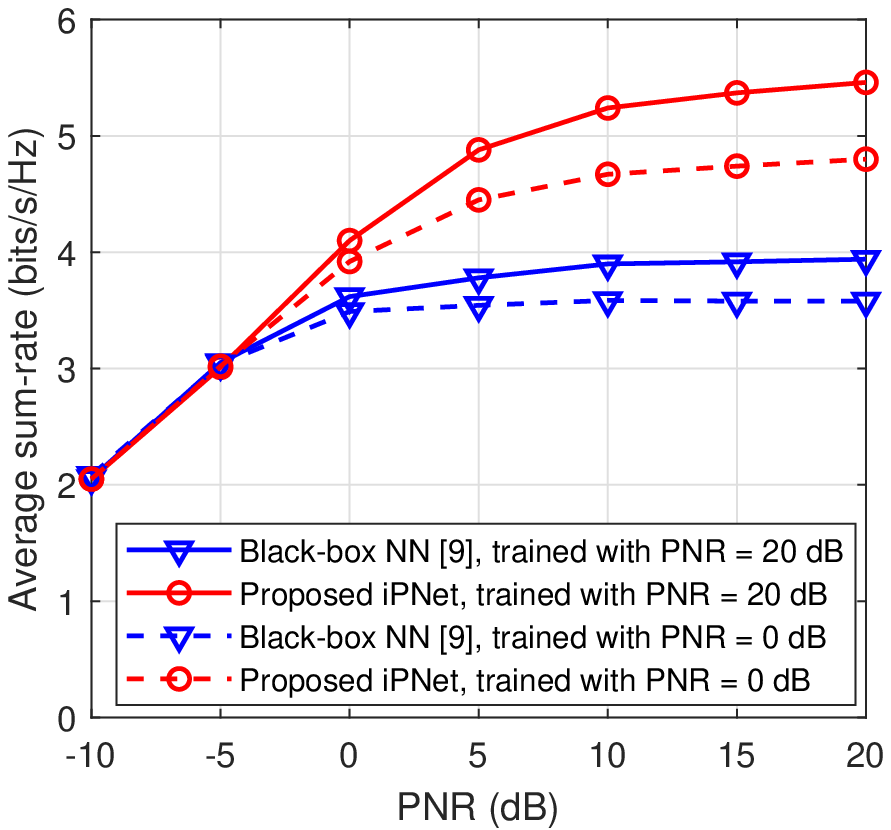}
\end{minipage}%
}
\subfigure[]{
\begin{minipage}[t]{0.5\linewidth}
\centering
\includegraphics[scale=0.6]{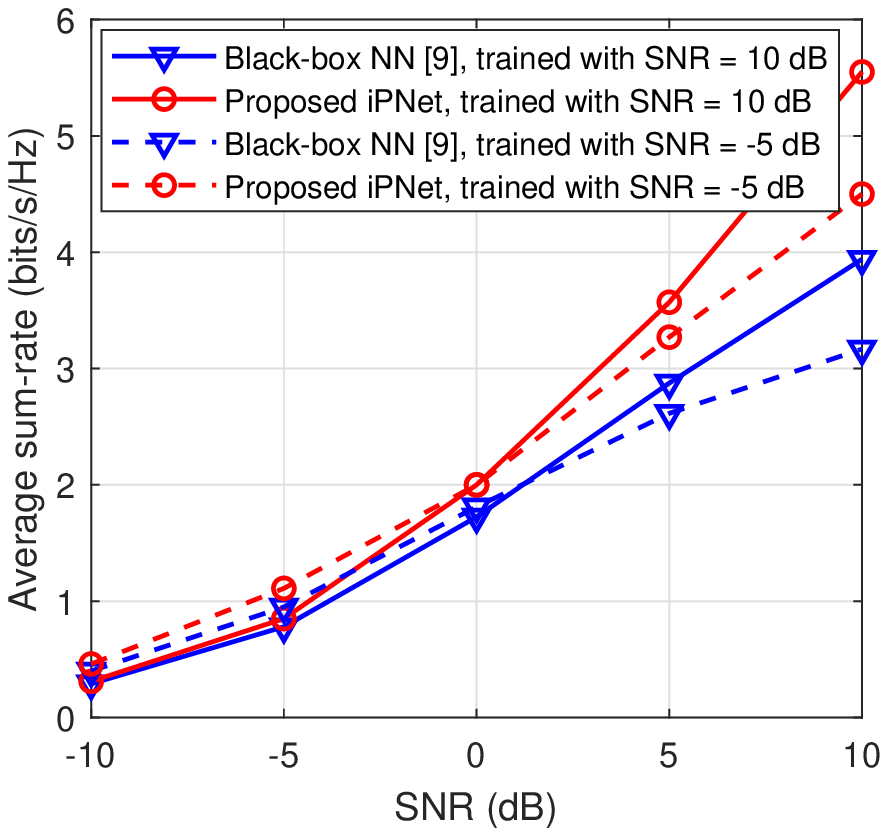}
\end{minipage}%
}%
\centering
\caption{(a). Generalization performance versus PNR. (b). Generalization performance versus SNR.}
\end{figure}

\begin{figure}[t]
\centering
\subfigure[]{
\begin{minipage}[t]{0.5\linewidth}
\centering
\includegraphics[scale=0.6]{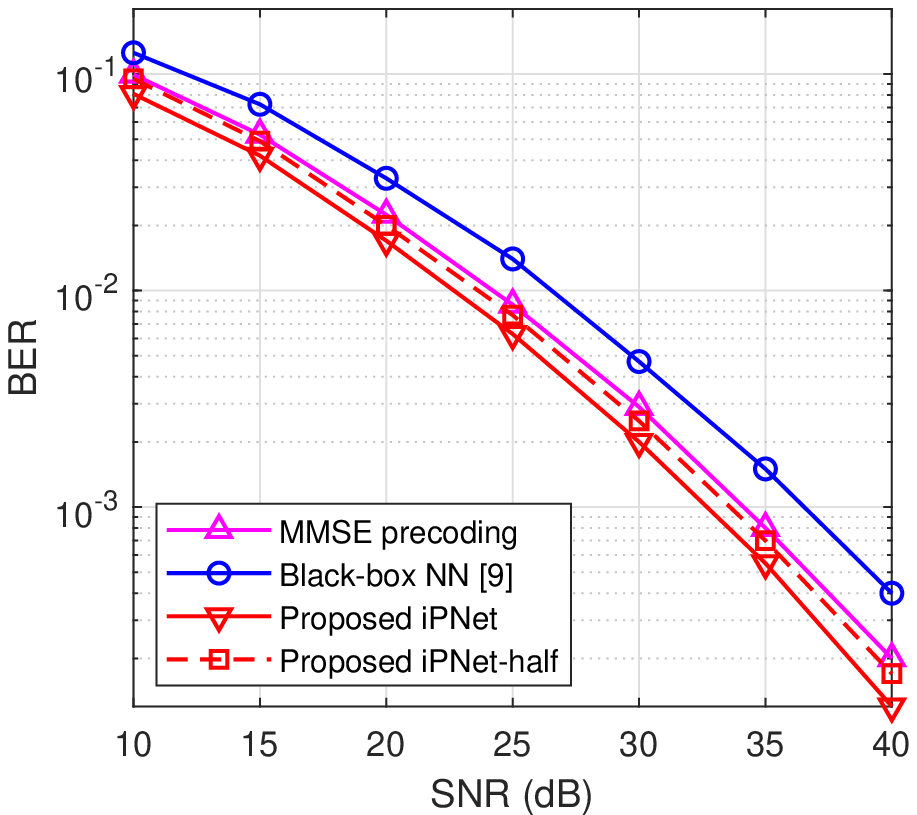}
\end{minipage}%
}
\subfigure[]{
\begin{minipage}[t]{0.5\linewidth}
\centering
\includegraphics[scale=0.6]{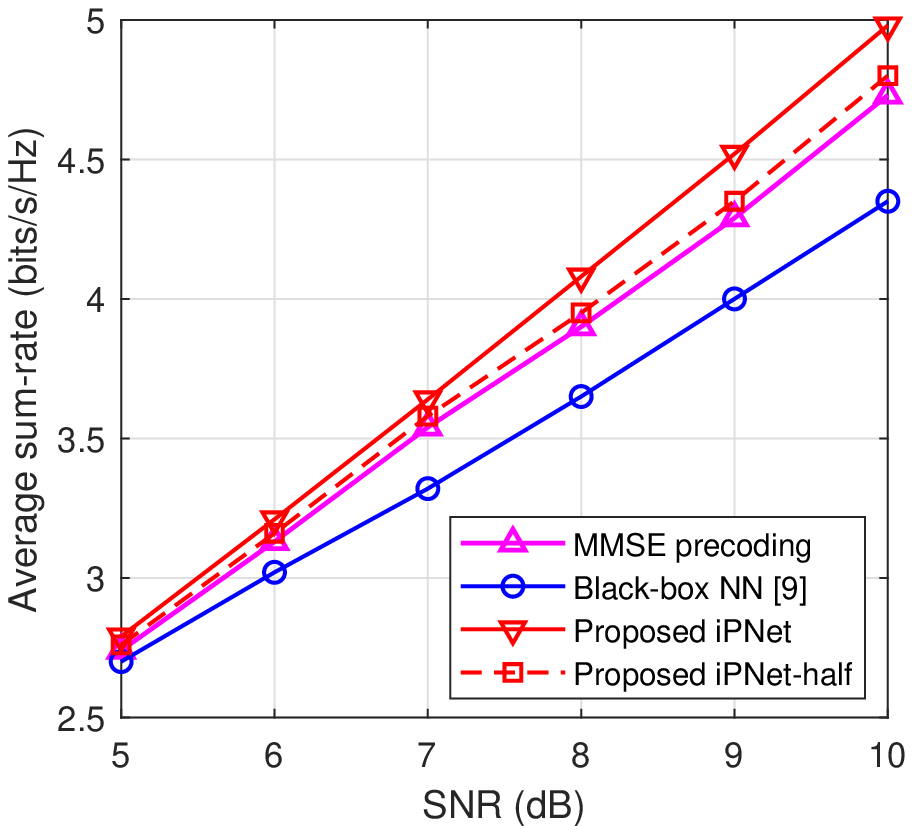}
\end{minipage}%
}%
\caption{(a). BER comparison. (b). Performance comparison of multiuser multiple-receiving-antenna single-stream scenario.}
\centering
\end{figure}


In Fig. 3, we compare the generalization performance for iPNet and the black-box NN, i.e., the robustness against imperfect CSI of a network trained under a fixed setup of PNR value or SNR value. In general, the imperfect CSI is obtained through the channel estimation. In Fig. 3(a), for the solid lines, the iPNet and the black-box NN are trained under the estimated CSI obtained at PNR = 20 dB. As for the dotted lines, they are trained under PNR = 0 dB. When the trained networks are tested with other PNR values in this figure, i.e., the estimated CSI obtained from other PNR values is input to the trained networks, it is observed that the proposed iPNet always outperforms the black-box NN and the performance gap increases with PNR. On the other hand, similar conclusions can be drawn from Fig. 3(b) when we observe the system performance versus SNR. These show the robustness of the proposed iPNet against imperfect CSI.
In particular, the enhanced ability of the proposed iPNet comes from the fact that the precoder network can learn from the augmented CSI data, rather than the raw CSI.

In Fig. 4(a), we compare bit error rate (BER) among these precoding methods. In particular, the Quadrature Phase Shift Keying (QPSK) modulation is adopted. It is observed that the proposed iPNet outperforms the MMSE precoding and the DL-based precoding method of [9] under different SNRs, which further validates the effectiveness of the proposed precoding method. On the other hand, in Fig. 4(b), we extend the proposed method to a multiuser single-stream system with a 4-antenna BS and two 2-antenna users. In particular, we fix the receiving combiner by using the maximum eigenvector of the corresponding channel matrix. 
The corresponding simulation results are shown in Fig. 4(b). We observe that the proposed iPNet outperforms the MMSE precoding and the DL-based precoding method of [9], which shows the effectiveness of the proposed precoding method for the extension to the scenario of multiuser multiple-antenna single-stream system. Note that for multi-antenna and multi-stream systems, iterative precoding algorithms that alternatingly optimize the precoding matrix jointly with the receiving weights have been developed, e.g., the WMMSE algorithm [1]. Yet, due to their iterative nature, the iPNet cannot be extended directly, therefore investigating DL-based methods with CSI augmentation to mimic the iterative precoding algorithms is left as an interesting future work.

\section{Conclusion}
This paper proposed a hybrid network composed of a model-driven sub-network and a data-driven sub-network for an interpretable neural precoding network design, namely iPNet. Due to the preprocessing of the model-driven network for input data augmentation, the data-driven sub-network becomes interpretable with stronger ability of generalization. Simulation results showed that the proposed iPNet performs better under various test conditions over various baseline schemes.

\bibliographystyle{IEEEtran} 
\bibliography{bare_jrnl.bib} 

\end{document}